\documentclass[9pt,twocolumn,twoside]{opticajnl}
\journal{opticajournal} 

\setboolean{shortarticle}{true}

\usepackage{comment}
\usepackage{lineno}
\usepackage{siunitx}
\usepackage{multirow}
\usepackage{ulem}
\usepackage{upgreek}

\title{Sub-kelvin measurement of silicon thermal expansion with a Fabry-P\'erot cavity stabilized laser} 

\author[1]{Pierre Roset}
\author[1]{Yara Hariri}
\author[1]{Rémi Meyer}
\author[1]{Jacques Millo}
\author[1]{Clément Lacroûte}
\author[1]{Samuel Margueron}
\author[1]{Yann Kersalé}
\author[1,*]{Jonathan Gillot}

\affil{Université Marie et Louis Pasteur, SUPMICROTECH, CNRS, institut FEMTO-ST, F-25000 Besançon, France}
\affil[*]{jonathan.gillot@femto-st.fr}

\begin{abstract}

In this letter, we report the measurement of the coefficient of thermal expansion (CTE) of single-crystal silicon from $\qty{655}{\milli\kelvin}$ to $\qty{16}{\kelvin}$ using an ultra-stable laser based on a single-crystal silicon Fabry-Perot cavity. Below $\qty{1}{\kelvin}$ temperatures, the CTE is in the $10^{-13}\  \unit{\per\kelvin}$ range with a lowest point at $\boldsymbol{\alpha(}\qty{655}{\milli\kelvin}\!\boldsymbol{)=}\qty{3.5\pm0.4 e-13}{\per\kelvin}$. We produce a theoretical model based on Debye and Einstein models to effectively approximate the CTE measured in this temperature range. 
This is the lowest-temperature CTE measurement of silicon to date, as well as the lowest operating temperature for an ultra-stable Fabry-Perot cavity for laser frequency stabilization.

\end{abstract}

\setboolean{displaycopyright}{false} 

\begin{document}

\maketitle

\section{Introduction}


Fabry-Perot cavities are optical resonators widely used in many fields of modern science. The ability to reduce its fractional length fluctuations, equal to fractional frequency fluctuations of the resonant mode, is particularly essential for optical frequency metrology \cite{salomon_laser_1988,ludlow_optical_2015,nicholson_systematic_2015} which also opens the door to fundamental physics tests \cite{Hild_2011,savalle_searching_2021}. The fundamental limit of the cavity length stability is the thermal Brownian noise arising from cavity constituents such as the mirrors (substrates and coatings) as well as the spacer that rigidly maintain mirrors in position \cite{kessler_thermal_2012}. Longer cavities \cite{parke_three_2025}, enlarging the laser spot size by increasing the radius of curvature (ROC) of the mirrors \cite{davila-rodriguez_compact_2017} or using high order Hermite-Gaussian modes \cite{zeng_thermal-noise-limited_2018} are solutions tested to over-come this limitation. Another possibility is to reduce the cavity thermal noise level directly by using a material with lower mechanical losses or by decreasing the operating temperature~\cite{kessler_thermal_2012}.

In the last decade, many developments of ultra-stable cavities have been conducted using low-loss materials at cryogenic temperatures, namely sapphire \cite{he_ultra-stable_2023} and crystalline silicon \cite{matei_15_2017}. With such cavities, state-of-the-art ultra-stable lasers (USL) with fractional frequency instabilities down to $4\times10^{-17}$ have been reported \cite{matei_15_2017}. Silicon has the major advantage to exhibit two cancellation points of its coefficient of thermal expansion close to \qty{17}{\kelvin} and \qty{124}{\kelvin}. Associated with a low noise temperature regulation, temperature-induced length-changes of the spacer are minimized to allow predictable low frequency drifts \cite{wiens_simplified_2020}. 
These performances make ultra-stable cryogenic silicon cavities a good candidates as a flywheel oscillators, a key element of optical timescales \cite{milner_demonstration_2019}. In this context, reducing the operating temperature further improves the cavities temperature sensitivity. 
Si CTE has been measured from 1.6 K to the melting point, but it remains poorly known under 6 K, although the existence of a third cancellation point below 3 K has been suggested  \cite{wiens_simplified_2020}. 

For that purpose, we designed a silicon cavity for operation in a dilution cryocooler \cite{barbarat_guidelines_2025}. With a laser frequency locked to this cavity, we have access to a measurement of the length variation of the spacer and therefore to the CTE of single-crystal silicon.

\begin{figure}[ht]
\centering
\includegraphics[width=0.95\linewidth]{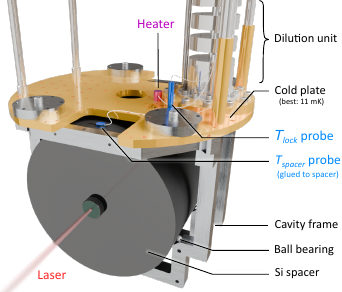}
\caption{3D model of the silicon cavity spacer, resting on three stainless steel ball bearings in a frame attached under the cold plate of the cryocooler. The two temperature RuO\textsubscript{2} probes are shown in light blue. The cold plate temperature is controlled using a dedicated heater (in purple) and monitored by the in-loop probe $T_{lock}$. Spacer temperature is monitored by an out-of-loop probe glued directly to the spacer ($T_{spacer}$).}
\label{fig:cavity_probes}
\end{figure}

\section{Experimental setup}

\subsection{Cavity properties}
The cavity consists of a spacer and two mirrors that are both made of single-crystalline silicon. The spacer is a \qty{18}{\cm}-long and \qty{20}{\cm}-diameter cylinder (Fig.~\ref{fig:cavity_probes}). The axis of the cylinder is aligned with the [111] crystalline orientation and bored for laser propagation. One flat and one concave (ROC of \qty{2}{\m}) mirrors are optically bonded to the spacer. The two mirror substrates are coated with a reflective crystalline GaAs/Al\textsubscript{0.92}Ga\textsubscript{0.08}As designed for cryogenic temperatures \cite{cole_high-performance_2016}.
We measured the cavity finesse by the ringdown method, obtaining $\mathcal{F}=420\,000 \pm 20\,000$ at low temperatures ($T_{spacer}<\qty{3}{\kelvin}$). The mode splitting, which is a birefringence effect observed in crystalline AlGaAs/GaAs coatings, has a value of \qty{267.309\pm0.001}{\kHz} at \qty{650}{\milli\kelvin}, consistent with a recent report \cite{yu_excess_2023}.

\subsection{Cryogenic system}
The cavity is cooled down to cryogenic temperatures with a commercial dilution cryocooler. The first cooling stage is provided by a pulse tube and allows to reach the temperature of \qty{3.6}{\kelvin}. A dilution stage, based on a mixture of $^3$He and $^4$He, cools down the coldest plate to \qty{11}{\milli\kelvin}.
At this temperature, the cooling capacity of the system is about \qty{10}{\micro\watt}, which roughly corresponds to the thermal radiation from windows. The cooling process from room temperature to \qty{11}{\milli\kelvin} takes 10 days and the lowest achieved cavity temperature is ${T}_{spacer}=\qty{585}{\milli\kelvin}$ (see Fig.~\ref{fig:cavity_probes}).
We use two calibrated ruthenium oxide (RuO\textsubscript{2}) sensors for the temperature control purpose. The out-of-loop  temperature sensor (${T}_{spacer}$) is fixed directly on the spacer to ensure an accurate knowledge of the cavity temperature.
We chose a bi-component glue with a good thermal conductivity of \qty{1.4}{\watt \per \meter \per \kelvin} at \qty{240}{\kelvin}, 
and we checked that this conductivity remains acceptable in the dilution regime by applying a temperature step and measuring a lag time lower than 1~s between the spacer temperature and the laser frequency.
The temperature control is performed via a heater and the in-loop (${T}_{lock}$) temperature sensor. A low noise AC resistance bridge controller acquires the temperature and generates the correction signal. This plate is typically regulated at \qty{400}{\milli\kelvin} such that the cavity temperature is set to \qty{650}{\milli\kelvin} with fluctuations below \qty{1}{\milli\kelvin} at \qty{1}{\s}.

\subsection{Optical setup}
\begin{figure}[ht] 
\centering
\includegraphics[width=1\linewidth]{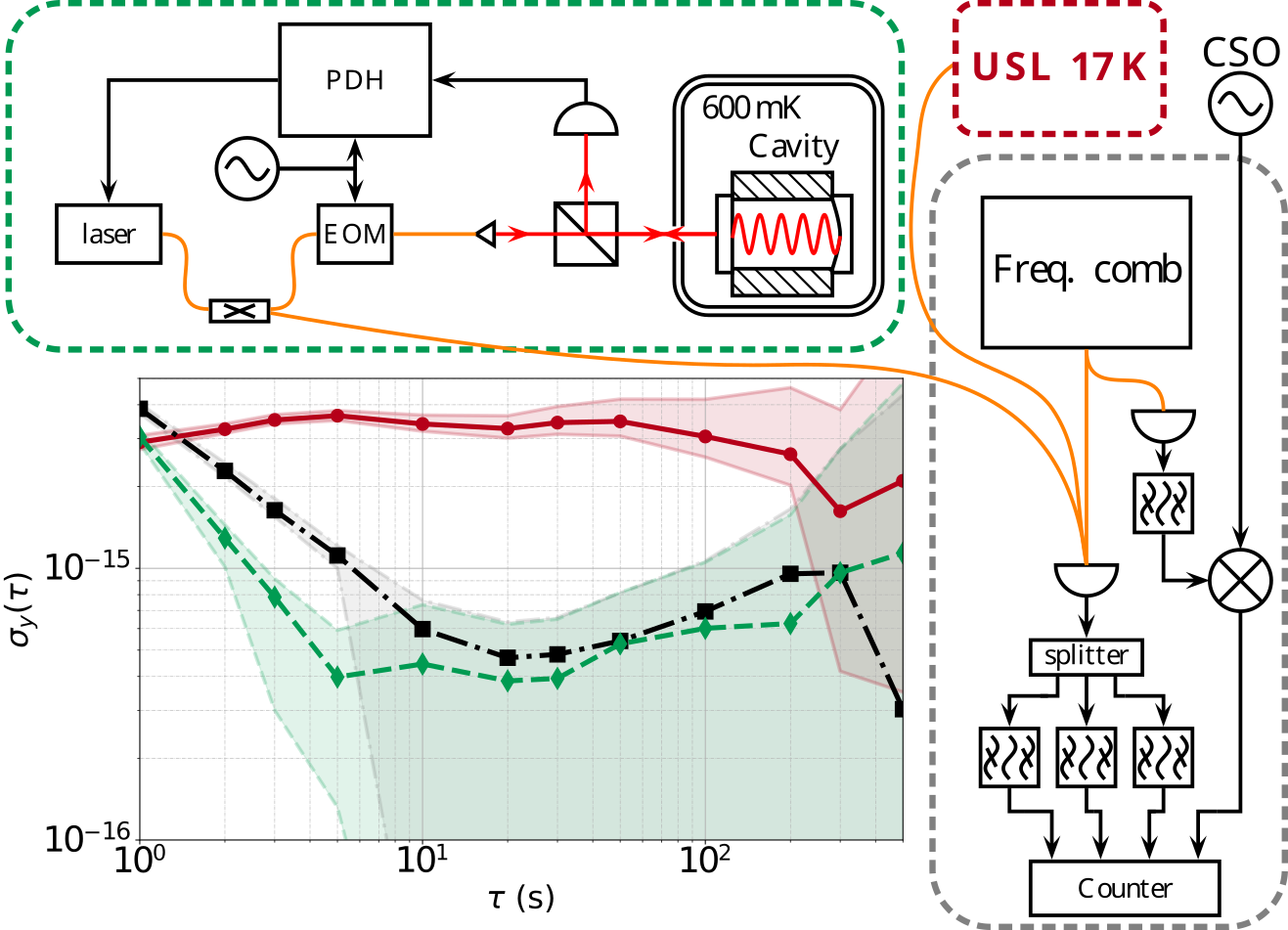}
\caption{Simplified scheme of the \qty{600}{\milli\kelvin} USL and the measurement setup. The dashed frames represent the different experiment involved; ultra stable signals are carried between those experiments with noise-compensated fiber links. Orange and red lines: laser propagation respectively in fibers and free-space. Black lines: electronic signals. EOM: Electro-Optic Modulator. USL \qty{17}{\kelvin}: USL stabilized on a silicon cavity at \qty{17}{\kelvin}. Inset: Allan deviations of the \qty{600}{\milli\kelvin} USL (dashed green line with diamond marker), the \qty{17}{\kelvin} USL (solid red line with dots marker) and a \qty{10}{\giga\hertz} CSO up-converted to optic (doted black line with square marker) obtained using the three-cornered hat method.}
\label{fig:cavity}
\end{figure}

Fig. (\ref{fig:cavity}) shows the optical setup implemented for laser frequency stabilization to the cavity. We stabilize the frequency of an erbium doped fiber laser on the cavity with the Pound-Drever-Hall (PDH) method \cite{salomon_laser_1988}. 
For that purpose, we use an electro-optic modulator (EOM) to modulate the laser electric field in phase and further generate a PDH error signal by using digital electronics. The mode matching and the beam injection through the plano mirror of the cavity are done in free space. A Faraday rotator placed in front of the cavity allows to select one of the birefringent modes. 
The laser frequency is corrected with a fibered acousto-optic modulator (AOM) for fast corrections and with the piezoelectric transducer of the laser for slower corrections, but with a larger dynamic. 
Since intra-cavity laser power fluctuations indirectly induce frequency fluctuations, we also implement a laser power stabilization through the RF power of the AOM \cite{tricot_power_2018}.
The optical power lock is achieved using analog electronics and the injected power is locked at \qty{300}{\nano\watt}.

The ultra-stable signal is transferred though a \qty{30}{\meter}  compensated fiber link \cite{mukherjee_link_2022} to another room in which means of comparison are available, including: (\textit{i}) a \qty{10}{\giga\hertz} signal provided by a cryogenic sapphire oscillator (CSO) \cite{fluhr_cryogenic_2023}, (\textit{ii}) an USL stabilized to a \qty{17}{\kelvin} silicon Fabry-Perot cavity  \cite{hariri_development_2024} and (\textit{iii}) a femtosecond laser providing an auto-referenced optical frequency comb.
We use these two references and the frequency comb to measure the fractional frequency stability of the laser stabilized to the \qty{600}{\milli\kelvin} cavity through the 3-cornered hat measurement. The frequency comb is used to multiply the microwave CSO signal to optical frequencies. The laser stabilized to the \qty{17}{\kelvin} cavity and the CSO have fractional frequency stabilities of respectively $3\times 10^{-15}$ before \qty{100}{\s}, and \num{6e-16} between \qty{10}{\s} and \qty{100}{\s}. For the laser stabilized to the sub-K cavity, we report an instability below \num{1e-15} between \qty{2}{\s} and \qty{200}{\s} and close to \num{4e-16} between \qty{5}{\s} and \qty{30}{\s} (see inset Fig. \ref{fig:cavity}).
Although the expected cavity thermal noise is $\simeq\num{3e-18}$ at \qty{600}{\milli\kelvin}, these performances are more than sufficient to measure the CTE of silicon.

\section{Coefficient of thermal expansion}

\subsection{Measurement methods}

\begin{figure}[t]
\centering
\includegraphics[width=1\linewidth]{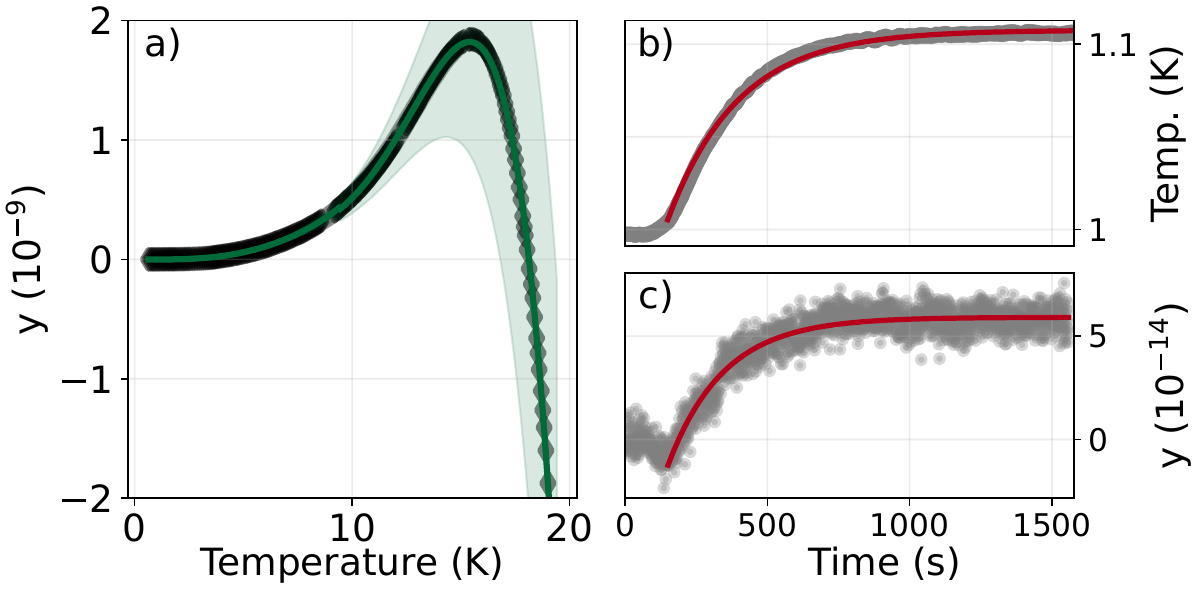}
\caption{a): fractional frequency variation (y$=\Delta\nu/\nu_0$) versus temperature for a dynamical measurement from \qtyrange[range-units = single]{0.6}{18}{\kelvin}. The treated data are represented by black dots. The primitives of the CTE $5^\text{th}$-order polynomial fits (\eqref{eq:fitpoly}) are also represented (plain green line). b):~Cavity temperature ($T_{spacer}$ on Fig. \ref{fig:cavity_probes}) response to a temperature step as a function of time. c):~Laser frequency response to a temperature step as a function of time. Gray dots show the raw data in plot b) and c), fitted  with \eqref{expfit} (plain red curve) to extract the amplitude of the step response. }
\label{fig:alpha_threatement}
\end{figure}

Two methods were used to extract the silicon CTE (Figure \ref{fig:alpha_threatement}).
The first one is a so-called "dynamic" measurement. Starting at the lowest temperature, we induce a large temperature sweep and we record the cavity temperature and the laser frequency averaged with a gate time of 5 s.  
The temperature dataset is split into subsets of 100 mK. The same pattern is used to also split the frequency dataset into subsets. 
To reject the short-term noise, the temperature and frequency of each subset are averaged. The CTE $\alpha$ is given by:
\begin{equation}
    \alpha(T)=\frac{d\nu}{\nu_0}\frac{1}{dT},
\end{equation}
with $d\nu,dT$ respectively the frequency and temperature variations obtained from the treated data and $\nu_0$ the laser frequency.

The second method is based on step measurements of the CTE. It consists of measuring $\alpha$ after applying a small temperature step (on $T_{lock}$) and simultaneously recording $T_{spacer}$ and the laser frequency response until the stationary regime is reached.
We fit the recorded temperature and laser frequency with an exponential function (Fig. \ref{fig:alpha_threatement} b) and c)): \begin{equation}
   f(t;A,\tau,C)= Ae^{-\frac{t}{\tau}}+C.\label{expfit}
\end{equation}
The parameter $A$ represents the amplitude of the response of the laser frequency $\Delta \nu$ or temperature $\Delta T$ and
$\alpha$ is then computed with :
\begin{equation}
    \alpha(T)=\frac{\Delta \nu}{\nu_0}\frac{1}{\Delta T}.
\end{equation}
Contrary to the previous method, each step only provides one value for CTE, at the average temperature of the step. This method is more accurate but slower and implies that the temperature must be stabilized at each step. Constrains in the cryocooler dilution process do not allow to use this method for temperatures between \qty{1.2}{\kelvin} and \qty{3.6}{\kelvin} or above \qty{10}{\kelvin}.
However, this step method is used at sub-\qty{1}{\kelvin} temperatures to confirm the dynamic measurement visible on Fig. \ref{fig:alpha_sub-K}.

\subsection{Silicon CTE at low temperature}

\begin{figure}[t]
\centering
\includegraphics[width=1\linewidth]{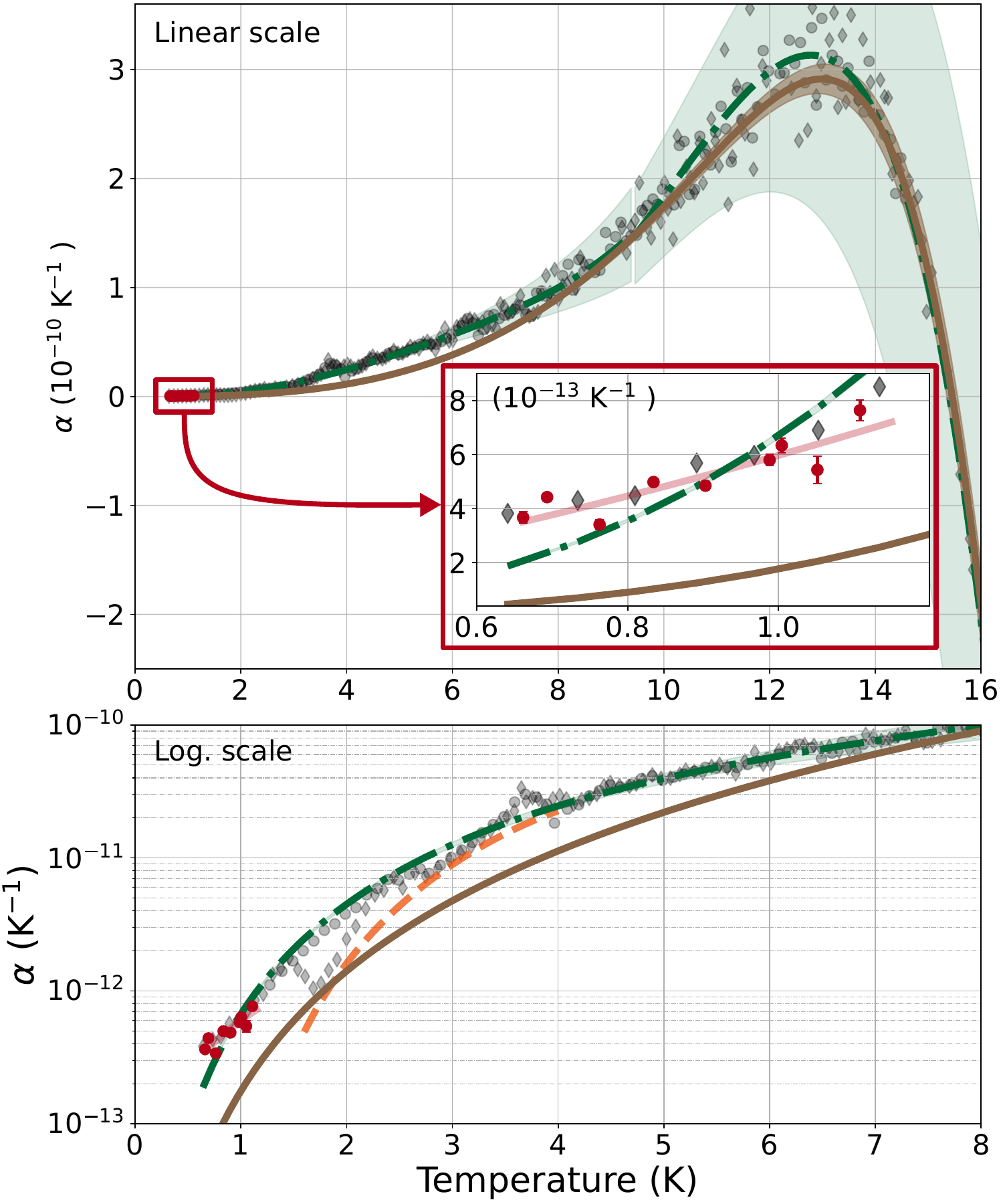}
\caption{CTE of Si measurements for temperatures between \qty{600}{\milli\kelvin} and \qty{16}{\kelvin}. Top: linear scale. Inset: zoom on the \qty{600}{\milli\kelvin} to \qty{1.15}{\kelvin} range. Bottom: zoom on the \qty{600}{\milli\kelvin} to \qty{8}{\kelvin} range in semi-log scale. Three datasets are shown, two dynamic sets (grey dots and diamonds) and a set of step measurements (red circles). Four different fits are used: green dashed curve, $5^\text{th}$-order polynomial fits (\eqref{eq:fitpoly}); plain brown curve, Einstein-Debye model (\eqref{eq:ED_model}); orange dashed curve, $5^\text{th}$-order polynomial function used by \cite{wiens_optical_2023}; red plain curve, $2^\text{nd}$-order polynomial fit of the step data (\eqref{eq:lowpoly}).}
\label{fig:alpha_sub-K}
\end{figure}

Figure \ref{fig:alpha_sub-K} shows the silicon CTE from \qty{0.6}{\kelvin} to \qty{16}{\kelvin}. We measure the second zero-crossing point of silicon CTE at \qty{15.45\pm0.1}{\kelvin}, which is in agreement with previously reported values \cite{middelmann_thermal_2015,lyon_linear_1977}. The repeatability of the measurement is shown by the superposition of two dynamic datasets (gray dots and diamonds). The step measurements is restrained to temperatures below \qty{1.2}{\kelvin}. They are in a excellent agreement with the dynamic data (red circles, see inset Fig. \ref{fig:alpha_sub-K}). 

Below \qty{1.2}{\kelvin}, the CTE is below \qty{1e-12}{\per\kelvin} with the following 2$^\text{nd}$-order polynomial fit describing its behavior: 
\begin{equation}
    \alpha_{\qtyrange[range-phrase = -,range-units = single]{0.6}{1.15}{\kelvin}}(T)= \num{4\pm 1.6 e-13}T+ \num{2\pm 1.7 e-13}T^2\label{eq:lowpoly}
\end{equation}

\section{Discussion}

In most of publications about the Si CTE related to ultra-stable cavities, the measured data is described using polynomials~\cite{wiens_optical_2023,wiens_simplified_2020,lyon_linear_1977}. This approximation is valid for a restricted interval of temperature. For larger intervals, higher order polynomial is needed and more instabilities are induced by the fit (mainly oscillations on the fit edges). To limit these instabilities, we use two 5$^\text{th}$-order polynomial fits \eqref{eq:fitpoly}, one between \qtylist[list-units = single]{0.6;10.2}{\kelvin} and the second between \qtylist[list-units = single]{10.2;16}{\kelvin}. The zero and first-order coefficients are set to 0 because the CTE and its tangent are null at \qty{0}{\kelvin}. The second-order is also set to 0 as it induces coupling between the different coefficients, reducing the accuracy of the fit. This fit uses the same layout than the one used by \cite{wiens_optical_2023}: 
\begin{equation}
    \alpha_\text{poly}(T;c_3,c_4,c_5)=c_3T^3+c_4T^4+c_5T^5\label{eq:fitpoly}.
\end{equation}
For sub-\qty{1}{\kelvin} temperatures, the CTE exhibits values in the mid $10^{-13}~\unit{\per\kelvin}$.
A slight difference between our polynomial fit and \cite{wiens_optical_2023} is visible on Fig. \ref{fig:alpha_sub-K}. Considering that the CTE is highly dependent on the experimental system, this difference is however very small (at most $10^{-11} \unit{\per\kelvin}$) .\\

\begin{table}[ht]
    \centering
     \begin{tabular}{|p{1.2cm}||p{2.8cm}|p{2.8cm}|}
         \hline
           & \qtyrange[range-units = single]{0.6}{10.2}{\kelvin} & \qtyrange[range-units = single]{10.2}{16}{\kelvin}\\
         \hline
         $c_3$ (\unit{\kelvin^{-4}}) & \num{-1.987 \pm 0.065 e-13} &\num{1.977 \pm 0.024 e-13}  \\
         $c_4$ (\unit{\kelvin^{-5}}) & \num{3.648 \pm 0.07 e-14} &\num{-2.575 \pm 0.055 e-14}  \\
         $c_5$ (\unit{\kelvin^{-6}}) &\num{-1.415 \pm 0.019 e-15} &\num{1.129 \pm 0.031 e-15}  \\
         \hline
     \end{tabular}
    \caption{Coefficients of the $5^\text{th}$-order polynomial functions \eqref{eq:fitpoly}}.
    \label{tab:coef}
\end{table} 
The specific heat $c_v(T)$ and the phononic density of states are directly related to the CTE. The dispersion is composed of three optical, two transverse (TA) and one longitudinal (LA) modes because silicon presents a face-centered cubic crystal with 2 atoms per cell (6 modes are possible). Nevertheless, the three optical modes do not contribute much at low temperature so they will be neglected here. Middelmann \textit{et al.} approximate the contribution to these 3 modes (2 TA and 1 LA) for heat capacity with Einstein's model \cite{middelmann_thermal_2015}, such as
\begin{equation}
    \alpha(T)=\frac{1}{3B}\sum_i \gamma_i c_v^i(T),\label{eq.alpha_middl}
\end{equation}
with $\gamma_i$ the Grüneisen coefficients describing the positive or negative contribution of phonons with respect to the volume expansion, and $B$ the bulk modulus.\\ 
Meanwhile, observing the dispersion branch of silicon, one can classify these 3 modes in two groups: two TA modes with a cutoff frequency around \qty{4.2}{\tera\hertz} and one LA mode with a cut-off frequency of \qty{12.5}{\tera\hertz}. 
Table \ref{tab:coef} indicates that at low temperatures the CTE can be approximated by a Debye model with T$^3$ dependence. Finally, we propose an approximation of the silicon coefficient of thermal expansion with only two contributions: 
\begin{equation}
    \alpha(T)=\frac{1}{3}\left(\frac{2}{3B_\text{TA}}\gamma{_{_{\text{TA}}}} c_v^\text{E}(T)+ \frac{1}{3B_\text{LA}}\gamma{_{_{\text{LA}}}}c_v^\text{D}(T)\right),\label{eq:ED_model}
\end{equation}
where $B_\text{TA}=\qty{70} {\giga\pascal}$ and $B_\text{LA}=\qty{165} {\giga\pascal}$ are the transverse and longitudinal bulk modulus respectively \cite{holland_analysis_1963}, and $\gamma_{_\text{TA/LA}}$ the transverse or longitudinal mode Grüneisen coefficients. The Einstein model contribution to heat capacity can be written as
\begin{equation}
    c_v^\text{E}(T) = \frac{N_a k_B}{V_m} \left( \frac{\theta_\text{E}}{T} \right)^2 \frac{e^{\theta_\text{E}/T}}{(e^{\theta_\text{E}/T}-1)^2},
\end{equation}
with $\theta_\text{E}=\qty{200}{\kelvin}$ the Einstein temperature of TA modes \cite{ibach_thermal_1969}, $N_a$ the Avogadro number, $V_m$ the molar volume of silicon and $k_B$ the Boltzmann constant.
On the other hand, the Debye model contribution is 
\begin{equation}
    c_v^\text{D}(T) = \frac{3 N_a k_B}{V_m} \left( \frac{T}{\theta_\text{D}} \right)^3 \int_0^{\theta_\text{D}/T} \frac{x^4e^{x}}{(e^{x}-1)^2} dx,
\end{equation}
with $\theta_\text{D}=\qty{570}{\kelvin}$ the Debye temperature of the LA mode approximated as independent of temperature in the range of 0.6 to 16~K.
This model can only be used at low temperatures, otherwise some phonon-phonon interactions need to be taken into account~\cite{holland_analysis_1963}.

The adjustment of this model of thermal expansion to our data gives the Gruneïsen coefficients $\gamma_\text{TA}=\num{-0.327 \pm 0.004}$ and $\gamma_\text{LA}=\num{0.902 \pm 0.028}$. It can be noted that those coefficients carry the sign of the CTE as stated in \cite{xu_theory_1991}. The Gruneïsen coefficient of TA modes is expected to be negative, contrarily to the LA modes, that participate to the positive part of the CTE.\\
\indent In Fig. \ref{fig:alpha_sub-K}, a slight discrepancy is visible between the Einstein-Debye model and the data at low temperatures ($T<\qty{7}{\kelvin}$). This may be explained by impurities in the spacer lattice causing disruptions in the density of phonon states of Si.

\section{Conclusion}
We reported the measurement of the Si CTE at unprecedented low temperatures between \qtylist[list-units = single]{0.655;16}{\kelvin}, with a laser frequency stabilized to a Fabry-Perot cavity. The lowest zero-crossing point of Si has been measured at \qty{15.45\pm0.1}{\kelvin} with a residual sensitivity in the $10^{-10}$~\unit{\per\kelvin\squared} range. At sub-\qty{1}{\kelvin} temperatures, the CTE is below \qty{1e-12}{\per\kelvin} and as low as $\alpha=\qty{3.5\pm0.4 e-13}{\per\kelvin}$ for 655 mK, the coldest temperature achieved. 
A theoretical model based on Debye and Einstein models has been proposed and shows a very good agreement with our measures between \qtylist[list-units = single]{0.6;16}{\kelvin}. We also demonstrate the possibility to operate an USL stabilized on a Fabry-Perot cavity at sub-\qty{1}{\kelvin} temperature with a fractional frequency stability in the $10^{-16}$ range.

\begin{backmatter}
\bmsection{Funding} This work has been supported by the ANR-17-EURE-0002 EIPHI Graduate School, the ANR-10-LABX-48-01 First-TF and ANR-11-EQPX-0033 Oscillator~IMP and the Région Bourgogne Franche-Comté.

\bmsection{Acknowledgment} The authors are indebted to Thomas Legero for helping with the optical bonding of the cavity mirrors. We would like to thank Vincent Giordano and Christophe Fluhr for fruitful advices on thermal design and cryocooler operation. We thank Philippe Abbé for his technical support and general expertise, and Rodolphe Boudot and Marion Delehaye for their useful suggestions.

\end{backmatter}

\bibliography{article_optica}

\bibliographyfullrefs{article_optica}

\end{document}